\documentclass[prb,aps,twocolumn,showpacs]{revtex4-1}%
\usepackage{amsmath}
\usepackage{amsfonts}
\usepackage{amssymb}
\usepackage{graphicx}%
\usepackage{bm}
\begin{document}

\title{Reply to Comment  by E. Babaev and M.
Silaev, arXiv:1105.3756}
\author{ V. G. Kogan}
%\affiliation{Ames Laboratory DOE, Iowa State
%University, Ames, Iowa 50011, USA }
\author{J\"{o}rg Schmalian}
\affiliation{Department of Physics and Astronomy and Ames Laboratory, Iowa State
University, Ames, Iowa 50011, USA }
\date{\today}

\begin{abstract}
The criticism of Babaev and Silaev notwithstanding, we conclude that our analysis is correct. We have found in our papers on two-band superconductors close to Tc, where the Ginzburg-Landau (GL) theory applies, that these materials are characterized by a single order parameter, governed by a single correlation length. In the GL domain, the order parameters of individual bands are proportional to each other. This happens due to the unavoidable inter-band Josephson coupling. Consequently, in the regime where the GL theory applies, these systems are either type-I or type-II superconductors with no room for so called "1.5-type" superconductivity. This conclusion does not mean that at lower temperatures, outside of the GL domain, the inter-vortex interaction cannot have interesting properties, however, the latter cannot be addressed with the GL formalism.
\end{abstract}

 \pacs{}
\maketitle

In a recent Comment\cite{Comment} Babaev and Silaev (BS) state that our
work on coupled two band superconductors\cite{Geyer,Kogan11} is 
incorrect.  After considering the criticism we still conclude  
 that our analysis is correct. The
essence of our conclusion is that multiband superconductors close to $T_{c}$, where the Ginzburg-Landau (GL) theory applies, are characterized by a single
length scale of the order parameter change, or by a single correlation length. This is due to the
 inevitable interband Josephson coupling. To keep the Reply brief we respond to major  arguments  of the  Comment. 
 %that can be found in Ref.\,\onlinecite{Comment}.

%\textbf{Systems with }$U(1)\times U(1)$\textbf{\ symmetry:} \ 
1. {\it Systems with  $U(1)\times U(1)$  symmetry}. The Comment argues that
there are systems where the coupling between two condensates vanishes and
  therefore our theory doesn't apply. This might be  the case for
neutron-proton condensation in nuclei or for condensed electrons and protons in the hypothetic hydrogen metal. In our papers we made clear that
we are concerned with multiband superconductors  were the interband 
Josephson coupling cannot be avoided.
%, as BS acknowledge. 
%Our theory clearly does
%not apply to neutron-proton condensation. It is for the readers to decide
%whether our theory for multiband superconductors is incorrect as it only
%applies to superconductors and not to neutron-proton condensation in nuclei.
In this sense, the Comment misses  the point: we did not discuss such systems in our papers.

2. {\it On the definition of the GL regime}. What is stated in
our papers is equivalent to the assertion that the   free energy expansion in powers of the order parameter  
 is only valid near the transition temperature, where the
order parameter is small, the original idea of Landau.  We stressed  that we ignore critical
fluctuations (as do BS in their respective papers). With this assumption the
restriction of the GL theory to the vicinity of $T_{c}$ is
hardly a debatable issue. The objections of the Comment are not quite consistent: one the one hand, BS claim that we cannot consider the limit $T\to T_c$ due to critical fluctuations and, on the other, they employ the mean-field GL functional without terms responsible for fluctuations. Both BS and ourselves disregard critical fluctuations, which is well justified for conventional superconductors.  

3. {\it  The GL domain and length scales.} As we discuss in our paper, the size of the domain where the GL equations hold varies from one system to the other. E.g., in the limit of weak
Josephson interband coupling   the order parameter of one band increases  quickly compared to the other.\cite{Geyer}  Therefore, one can only
perform the GL expansion in a very narrow region around $T_{c}$ and microscopic corrections to GL become important quickly. Another example of a very narrow GL domain is given in Ref.\,\onlinecite{Kosh}.
But however narrow the GL domain is, as long as we are in this domain and use the GL theory within its accuracy, both order parameters vary on the same length scale, the main result of our paper. The statement of two separate length scales {\it within the standard GL framework} is unsustainable. One might consider higher order terms in the GL  expansion\cite{Milos}   and explore the possibility of two different length scales, but their difference is going to be of a higher order in $1-T/T_c$ than the GL theory allows. 

One of course can formally employ 
GL functionals outside  the GL region but this does not make  much   sense. 
Instead one has to use   a microscopic theory, such as the Gor'kov or   
Bogoliubov-de-Gennes weak coupling theories. 
 In our view, looking for the potentially interesting  physics of intervortex interactions within GL formalism will not be fruitful. In the regime where GL applies, one only finds   one characteristic length scale for the order parameter change, a conclusion that is
hardly surprising. 
In this sense, the recent attempt by the authors  of the Comment to use  microscopic theory is welcome,\cite{micro} but in the GL domain it should confirm the GL conclusion of a single coherence length  as $T\to T_c$.\cite{Milos} 

And the last but not least: to our knowledge, there is no experimental evidence for a 1.5-type of behavior in any known superconductor. The clustering reported in Ref.\,\onlinecite{Mosch} on decorations of vortices in extremely small fields (with the average intervortex spacing of a few London penetration depths) may have a more prosaic sample inhomogeneity as a source. Attempts to see the two-length scales in the vortex core structure of MgB$_2$ have been unsuccessful.\cite{Morten} We conclude that 1.5-type superconductivity remains a speculation questionable theoretically 
%(not counting an item in Wikipedia on Type-1.5 superconductors) 
and unconfirmed  by experiment.


\begin{thebibliography}{9}

\bibitem{Comment} E. Babaev, M. Silaev, arXiv:1105.3756

\bibitem{Geyer} J. Geyer, R. M. Fernandes, V. G. Kogan, J. Schmalian, Phys.
Rev. B \textbf{82}, 104521 (2010).

\bibitem{Kogan11} V. G. Kogan and J. Schmalian, Phys. Rev. B \textbf{83},
054515 (2011).

\bibitem{Kosh} A. E. Koshelev and A. A. Golubov, \prl {\bf 92}, 107008  (2004).  

\bibitem{Milos}A. A. Shanenko, M. V. Miloÿsevi«c, F. M. Peeters, and A. V. Vagov, \prl  {\bf 106}, 047005 (2011).
 

\bibitem{Mosch}V. Moschalkov {\it et al}, 
%M. Menghini, T. Nishio,  Q. H. Chen,  A.V. Silhanek,  V. H. Dao, 
%L. F. Chibotaru,  N. D. Zhigadlo,  and J. Karpinski, 
\prl {\bf 102}, 117001 (2009).

\bibitem{micro}M. Silaev, E. Babaev, arXiv:1105.5734.

\bibitem{Morten}M. Eskildsen {\it et al}, Physica C {\bf 385}(1-2), 169 (2003). 

%\bibitem{wiki} .
\end{thebibliography}
\end{document}